\documentclass[12pt,final]{elsart}

  \usepackage{latexsym,bm,amsmath,amssymb,amsfonts}
  \usepackage{epsfig,graphics,graphicx}
  \usepackage{slashed}

  \long\def\comment#1{ }

  \newcommand{\beq}{\begin{eqnarray}}
  \newcommand{\eeq}{\end{eqnarray}}
  
 \def\simge{\mathrel{%
   \rlap{\raise 0.511ex \hbox{$>$}}{\lower 0.511ex \hbox{$\sim$}}}}
\def\simle{\mathrel{
   \rlap{\raise 0.511ex \hbox{$<$}}{\lower 0.511ex \hbox{$\sim$}}}}

\begin{document}

\begin{frontmatter}

\parbox[]{16.0cm}{ \begin{center}
\title{Gluon polarization in the nucleon demystified}

\author{Yoshitaka Hatta  }

\address{  Graduate School of Pure and Applied Sciences, University
of Tsukuba, \\Tsukuba, Ibaraki 305-8571, Japan
}



\begin{abstract}
Recently, X.~Chen {\it et al.} proposed a new approach to the 
 gauge invariant decomposition of the nucleon spin  into 
helicity and orbital parts.  
The key ingredient in their construction is the separation of the gauge field into ``physical" and ``pure gauge" parts.
We suggest a simple separation scheme and show that 
the resulting gluon helicity  
 coincides with the first moment of the conventional  polarized gluon distribution measurable in high energy 
experiments.

\end{abstract}
\end{center}}

\end{frontmatter}

\vspace{10mm}

Despite its intuitive clarity, the decomposition of the nucleon spin into the helicity and orbital angular momentum of quarks and gluons has remained one of the most elusive problems in QCD spin physics \cite{Jaffe:1989jz}.   The current unsatisfactory situation may be epitomized  by the following dilemma:   
On one hand, continuous efforts have been made both at experimental facilities and by the theorists to assess the gluon helicity contribution $\Delta G$ \cite{Aidala:2010af} defined as the first moment of the polarized gluon distribution. On the other hand, 
 since the seminal work of Ji \cite{Ji:1996ek}, it has been widely recognized by the community that 
 the gluonic angular momentum cannot be decomposed into helicity and orbital parts in a gauge invariant way. 
 This implies that  $\Delta G$ extracted from measurements and global QCD analyses has unfortunately no natural counterpart in Ji's framework, making one wonder what exactly the physical meaning of $\Delta G$ is. The case of the gluon orbital angular momentum is even murkier since there is no known way of directly measuring it, nor is its operator representation available.
 
Recently, however, the situation took an interesting turn when Chen {\it et al.} proposed a 
new, complete decomposition of the nucleon spin \cite{Chen:2008ag,Chen:2009mr}. 
 The key ingredient in their construction is the separation of the total gauge field into ``physical" and ``pure gauge'' parts 
\beq
&& A^\mu=A^\mu_{phys}+A^\mu_{pure} \label{dec}\,, 
\nonumber \\ 
&& F_{pure}^{\mu\nu} 
= \partial^\mu A^\nu_{pure} -\partial^\nu A^\mu_{pure} +ig [A^\mu_{pure},A^\nu_{pure}] = 0\,, 
\label{dec}
\eeq         
 which transform differently under gauge transformations 
 \beq
 && A^\mu_{phys} \to U^\dagger A^\mu_{phys}U\,, \nonumber \\
 && A^\mu_{pure} \to U^\dagger A^\mu_{pure} U -\frac{i}{g} U^\dagger \partial^\mu U \label{tran}
 \eeq
   The separation (\ref{dec}) is not unique, and accordingly, the gluon helicity contribution is scheme dependent.  Chen {\it et al.}
    imposed a subsidiary condition that can be used to construct $A^\mu_{phys}$ perturbatively, and found that the corresponding gluon helicity indeed differs from $\Delta G$ \cite{Chen:2011gn}. [See, also, \cite{Cho:2010cw}.]  While their scheme has some attractive physical features, how to test their predictions in experiment is currently unknown.    The experimental observability is a prerequisite  for a good definition of the gluon helicity.  One then asks the question: ``Is there a separation scheme (\ref{dec}) in which the gluon helicity coincides with $\Delta G$?"
In this letter we answer this question positively in the hope that the conflicting opinions about the nature of the gluon helicity are  reconciled with each other. \\
         
The original proposal by Chen {\it et al.}  \cite{Chen:2008ag,Chen:2009mr} achieves a 
complete decomposition of the QCD angular momentum operator into quarks' and gluons' 
helicity and orbital angular momentum.  This was further elaborated by Wakamatsu 
\cite{Wakamatsu:2010qj,Wakamatsu:2010cb}  where the covariant generalization of the decomposition was derived. 
 The result is \cite{Wakamatsu:2010cb}
 \beq
 M_{quark-spin}^{\mu\nu\lambda} &=& \frac{1}{2}\epsilon^{\mu\nu\lambda\sigma}\bar{\psi} 
\gamma_5 \gamma_\sigma  \psi\,, \\ 
 M_{quark-orbit}^{\mu\nu\lambda}&=&\bar{\psi}\gamma^\mu (x^\nu iD^\lambda
-x^\lambda iD^\nu )\psi\,,  \label{24} \\  
 M_{gluon-spin}^{\mu\nu\lambda}&=&  F_a^{\mu\lambda}A_{phys}^{\nu a} -
F_a^{\mu\nu}A_{phys}^{\lambda a}  \,,  \label{glu}  \label{25} \\
 M_{gluon-orbit}^{\mu\nu\lambda}&=&  F_a^{\mu\alpha}\bigl(x^\nu (D_{pure}^\lambda A_\alpha^{phys})_a
-x^\lambda (D^\nu_{pure}A_\alpha^{phys})_a \bigr)  \nonumber \\ && \qquad \qquad \qquad \qquad 
+ (D_\alpha F^{\alpha\mu})_a 
(x^\nu A_{phys}^{\lambda a} -x^\lambda A_{phys}^{\nu a}) \,.   \label{26}
\eeq
where $D^\mu_{pure}\equiv \partial^\mu + ig A_{pure}^\mu$ and  
$a,b=1,2,\cdots,8$ are the color indices. 
 We use the convention $\epsilon_{0123}=+1$. The second term on the right hand side of (\ref{26}) 
is gauge--invariant on its own, and Chen {\it et al.} included it in
  the quark--orbital part. (This would result in the change $D^\nu \to D^\nu_{pure}$ 
  in (\ref{24}).) Following Wakamatsu \cite{Wakamatsu:2010qj},
 we have relocated it to the gluon--orbital part. With this modification the quark 
 part agrees with Ji's decomposition and can be extracted from GPD analyses \cite{Ji:1996ek}. 
  The decomposition of the gluon spin into the helicity (\ref{25})  and orbital (\ref{26}) parts 
   has been made possible at the cost of introducing nonlocality: In general, 
   $A_{phys}^\mu$ and $A_{pure}^\mu$ are nonlocally related to the total $A^\mu$.   
 
 Let us focus on the gluon helicity operator (\ref{glu}). 
 We go to the infinite momentum frame and use the light--cone coordinates $x^\pm = \frac{1}{\sqrt{2}}(x^0 \pm x^3)$.
  In the framework of Chen {\it et al.}, the gluon helicity fraction of the nucleon spin is given by the matrix 
  element of the $\mu\nu\lambda=+12$ tensor component
   \beq
    \frac{-1}{2P^+} \langle PS| F_a^{+\mu}(0) \epsilon^{+-}_{\ \ \ \ \mu\nu}
   A^{\nu a}_{phys}(0)|PS\rangle\,. \label{def1}
   \eeq

  On the other hand, the conventional and experimentally accessible gluon helicity  
   is given by the first moment of the polarized gluon distribution 
   (see, e.g., \cite{Manohar:1990kr,Kodaira:1998jn}) 
  \beq
  \Delta G &=&  \int_0^1 dx_B \Delta g(x_B) \nonumber \\ 
   &=&  \frac{1}{4P^+} 
  \int_{-\infty}^\infty dy^- \epsilon(y^-) \langle PS| 
  F_a^{+\mu}(0)
 \nonumber \\ 
 && \qquad \qquad \qquad \times P \exp\left(-ig\int^{0}_{y^-} A^+(y'^-) dy'^- \right)_{ab} 
\epsilon^{+-}_{\ \ \ \ \mu\nu} F_b^{+\nu}(y^-) |PS\rangle\,, \label{def2}
  \eeq
     where $x_B$ is the usual Bjorken variable and the Wilson line is in the adjoint representation. ($P$ denotes path ordering.) 
   
 If we insist that the two definitions (\ref{def1}) and (\ref{def2}) are equivalent,  
 we must have that,
  using the notation    
     $x^\mu=(x^-,\vec{x})$ with $\vec{x}=(x^+, x^1,x^2)$, 
   \beq
A^{\mu a}_{phys}(x) \! \overset{\mbox{\normalsize{??}}}{=} \!   -\frac{1}{2}\int_{-\infty}^\infty dy^- \epsilon(y^- - x^-) P
 \exp\left(-ig\int^{x^-}_{y^-} A^+(y'^-,\vec{x}) dy'^- \right)_{ab} \!\!
F^{+\mu}_b(y^-,\vec{x})\,. \label{new}
\eeq
  Does this identification make sense? The right hand side obeys the 
  gauge transformation law (\ref{tran}) as expected for $A_{phys}^\mu$, but it is far from obvious 
  that the difference $A^\mu_{pure}=A^\mu-A^\mu_{phys}$ is pure gauge. 
   Remarkably, however, there exists a special, but very simple
    scheme  of separation (\ref{dec}) in which (\ref{new}) 
    becomes an identity rather than a definition.\footnote{Wakamatsu \cite{Wakamatsu:2010cb} discussed  the equivalence of the matrix elements (\ref{def1}) and (\ref{def2}) in the  light--cone gauge. Here we intend to show (\ref{new}) as an operator identity in generic gauges.}

In order to find such a scheme, we first observe that (\ref{new}) immediately implies 
that 
\beq
A_{phys}^+=0\,. \label{if}
\eeq
  Thus we may write, writing fields as matrices in the adjoint representation,     
\beq
A^+= A^+_{pure} = -\frac{i}{g}VW\partial^+ (VW)^\dagger = 
\frac{i}{g} \partial^+ (VW) (VW)^\dagger\,, \label{can}
\eeq
where 
\beq
&&  V(x)= P\exp\left(-ig\int^{x^-}_{\pm \infty} A^+(x'^-,\vec{x}) dx'^- \right)\,, \nonumber \\
&&  W(\vec{x}) = P\exp\left(-ig\int^{\vec{x}}_{\infty \vec{n}} \vec{A}(\pm \infty,\vec{x}')\cdot d\vec{x}' 
\right)\,. \label{rot}
\eeq
Note that $W$ is evaluated at $x^- =\pm \infty$ 
where the plus (minus) sign corresponds to the choice $x^-=+ \infty$ ($x^- = -\infty$) in the 
lower limit of the integration in $V$.
 The  path to spatial  infinity (denoted as `$\vec{x}=\infty \vec{n}$' with 
 $\vec{n}$ being a constant vector)
is arbitrary assuming that the field strength vanishes at  $x^- = \pm \infty$. 
 
 Promoting (\ref{can}) to a four--dimensional relation, we define 
\beq
A^\mu_{pure}\equiv - \frac{i}{g}VW \partial^\mu (VW)^\dagger\,, \label{pu}
\eeq
 which guarantees that  $F^{\mu\nu}_{pure}=0$,  
and 
\beq
A^\mu_{phys} \equiv A^\mu- A^\mu_{pure}\,. \label{ph}
\eeq 
In order for $A_{pure}^\mu$ to transform according to (\ref{tran}) 
under gauge transformation, we require that
\beq
\lim_{\begin{subarray}{c} x^- = \pm \infty \\ \vec{x}\to \infty \vec{n}  \end{subarray}} 
\partial_\mu U(x^-,\vec{x})=0\,, \label{fix}
\eeq
 that is, we allow only for 
 global gauge rotations 
at $(x^-,\vec{x})=(\pm \infty, \infty \vec{n})$. For consistency, $A_{pure}^\mu$ should 
vanish there,
 and this is in fact already implied by (\ref{pu}). 
Except for this minor qualification, our separation scheme is independent of 
the gauge choice.  
 
 Still, it will be very convenient in the following to consider the light--cone gauge which has a special status in our scheme and which can be
  accessed by setting $U=VW$ (consistently with (\ref{fix})).   
 Denoting fields in the light--cone gauge with a tilde, we find 
\beq
 \tilde{A}_{phys}^\mu &=& (VW)^\dagger A^\mu_{phys} VW\,,
 \nonumber \\ 
 \tilde{A}^\mu_{pure} &=& 0\,,
\eeq
 so that $\tilde{A}^\mu = \tilde{A}_{phys}^\mu$ in this gauge. The residual
  ($x^-$--independent) gauge 
 symmetry in the light--cone gauge is essentially contained in $W(\vec{x})$. 
  It may seem more natural to let $\tilde{A}^\mu_{pure}$ carry these degrees of freedom. 
  However, 
   we have absorbed them in $\tilde{A}_{phys}^\mu$ for our purpose. 
   By using these degrees of freedom, one can  fix the boundary condition for $\tilde{A}^\mu=\tilde{A}^\mu_{phys}$ as 
   $x^- \to \pm \infty$.    \\

We are now ready to prove (\ref{new}).  The last factor can be written as, suppressing color 
indices,  
\beq
P \exp\left(-ig\int^{x^-}_{y^-} A^+(y'^-,\vec{x}) dy'^- \right) 
F^{+\mu}(y^-,\vec{x}) &=& VW(x)(VW)^{\dagger}(y^-,\vec{x})F^{+\mu}(y^-,\vec{x})
\nonumber \\ 
&=&VW(x)\tilde{F}^{+\mu}_{phys}(y^-,\vec{x})
\nonumber \\ 
&=& VW(x) \frac{\partial}{\partial y^-} \tilde{A}^\mu_{phys}(y^-,\vec{x})\,,
\eeq
  where in the second equality we have used the fact that 
  $\tilde{F}^{\mu\nu}=\tilde{F}^{\mu\nu}_{phys}$ in the light--cone gauge. 
  [Remember that matrices are in the adjoint representation.] 
  The right hand side of (\ref{new}) then becomes  
 \beq
  && -\frac{1}{2}\int dy^- \epsilon(y^- - x^-)VW(x) 
  \frac{\partial}{\partial y^-} \tilde{A}^\mu_{phys}(y^-,\vec{x}) \nonumber \\ 
  &&  = 
  VW(x) \tilde{A}_{phys}^\mu (x) -\frac{1}{2}VW(x) \left(\tilde{A}_{phys}^\mu (\infty,\vec{x})
  + \tilde{A}_{phys}^\mu (-\infty, \vec{x}) \right) \nonumber \\
  && = A_{phys}^\mu (x) -\frac{1}{2}VW(x) \left(\tilde{A}^\mu_{phys} (\infty,\vec{x})
  + \tilde{A}^\mu_{phys} (-\infty, \vec{x}) \right)\,, \label{last}
  \eeq
   where we integrated by parts. (\ref{last}) differs from $A^\mu_{phys}$ by the surface terms
   at $x^- = \pm \infty$. However, these surface terms can be consistently eliminated. To see this, suppose that (\ref{new}) is valid. Then 
   \beq
&&  A_{phys}^{\mu } (\infty,\vec{x})= \frac{1}{2}\int dy^-  P
 \exp\left(-ig\int^{\infty}_{y^-} A^+(y'^-,\vec{x}) dy'^- \right)
F^{+\mu} (y^-,\vec{x})\,,  \nonumber \\ 
&& A_{phys}^\mu  (-\infty,\vec{x})=  - \frac{1}{2}\int dy^-  P
 \exp\left(-ig\int^{-\infty}_{y^-} A^+(y'^-,\vec{x}) dy'^- \right)
F^{+\mu}(y^-,\vec{x})\,.
 \eeq
   Going to the light--cone gauge, we get 
   \beq
   \tilde{A}_{phys}^\mu (\infty,\vec{x}) &=& (VW)^\dagger (\infty,\vec{x})A_{phys}^\mu (\infty,\vec{x})  \nonumber \\ 
   &=& \frac{1}{2}W^\dagger(\vec{x})\int dy^-  P
 \exp\left(-ig\int^{-\infty}_{y^-} A^+(y'^-,\vec{x}) dy'^- \right)
F^{+\mu} (y^-,\vec{x})\,,  \nonumber \\ 
&=& -W^\dagger (\vec{x}) A_{phys}^\mu (-\infty,\vec{x})  \nonumber \\ 
&=& -\tilde{A}_{phys}^\mu (-\infty,\vec{x})\,, 
  \eeq
   where, for definiteness, we have chosen $x^- = -\infty$ as the lower limit of the integration in (\ref{rot}). (The other case $x^- = \infty$ is a trivial modification.) 
    Therefore the surface terms in (\ref{last}) cancel and this completes the proof of (\ref{new}). 
   
   Note that the cancellation we have just observed is nothing but the well--known
    antisymmetric boundary condition of the gauge field in the light--cone gauge. 
    Thus in the above proof we have implicitly chosen this boundary condition by 
    adjusting $W(\vec{x})$.
    This is in accordance with the sign function $\epsilon (y^-)$ in (\ref{def2}), or equivalently, 
    the principal value prescription for the $1/x_B$ pole in $\Delta g(x_B)$
   \beq
   \int_{-\infty}^\infty dx_B \, \mbox{p.v.}\left(\frac{1}{x_B}\right) e^{iy^- P^+ x_B} 
   = i\pi \epsilon(y^-)\,.
   \eeq
    Different prescriptions for the $1/x_B$ pole lead to different boundary conditions for 
     $\tilde{A}^\mu=\tilde{A}^\mu_{phys}$ at $x^- \to \pm \infty$, just like the prescription 
     for the $1/k^+$ pole of the gluon propagator in the light--cone gauge \cite{Belitsky:2002sm}. 
     It does not matter which prescription one uses, since 
      the difference is proportional to $\delta (x_B)$ and 
     vanishes under the assumption that the $x_B$--integral converges as $x_B\to 0$. 
     However, it does change the appearance of $\Delta G$.  Had we chosen 
     a different prescription, say, $1/(x_B-i\epsilon)$, 
     we would have obtained a formula  
     for $\Delta G$ similar to (\ref{def2}), but with the step function $2\theta(y^-)$ in place of the sign function 
     $\epsilon(y^-)$. In the light--cone gauge, this corresponds to 
     the advanced boundary condition $\tilde{A}^\mu(\infty,\vec{x})=0$. 
     For each different prescription, the surface terms will
      be different. But they always vanish under 
     the corresponding boundary condition.  \\

In conclusion, the gauge--invariant decomposition of the gluonic contribution to the nucleon spin into helicity and orbital parts is not possible if one restricts to {\it local} operators \cite{Ji:1996ek}. Once one allows for {\it nonlocal} operators, it becomes possible \cite{Chen:2008ag}.   We have shown that the traditional  definition of the gluon helicity, $\Delta G$, can be nicely accommodated in this latter approach, thereby dispelling any concerns about the physical meaning of $\Delta G$. After all, $\Delta G$ is  measurable, gauge invariant, and meets the criterion by Chen {\it et al.} for a proper definition of the gluon helicity in QCD.      

By using the explicit relation between $A^\mu_{phys}$ and $A^\mu$, one can write down the all--order expression for the gluon orbital angular momentum (\ref{26}) as well. While it is measurable  as the difference between the total gluon contribution (from the GPD) and $\Delta G$, more direct access to the orbital component would of course be desirable.  \\

  {\it Acknowledgments}---I am grateful to Masashi Wakamatsu for keeping me interested in this problem through his works, and 
for valuable discussions.  I also thank Kazuhiro Tanaka for helpful correspondence.  
 This work is supported by Special Coordination Funds for Promoting Science and Technology 
 of the Ministry of Education, the Japanese Government.
\appendix

\end{document}